\documentclass[10pt,twocolumn,letterpaper]{article}

\usepackage{wacv}
\usepackage{times}
\usepackage{epsfig}
\usepackage{graphicx}
\usepackage{amsmath}
\usepackage{amssymb}
\usepackage{booktabs}
\usepackage{caption}
\usepackage[inline]{enumitem}
\usepackage{xspace}
\usepackage{comment}
\usepackage{caption}
\usepackage{subcaption}
\def\wacvPaperID{865} % Enter the WACV Paper ID here

\wacvalgorithmstrack   % Uncomment this line if you are submitting to the Algorithms Track.
\wacvfinalcopy % *** Uncomment this line for the final submission

\usepackage[accsupp]{axessibility}
\ifwacvfinal
\usepackage[breaklinks=true,bookmarks=false]{hyperref}
\else
\usepackage[pagebackref=true,breaklinks=true,colorlinks,bookmarks=false]{hyperref}
\fi

\begin{document}

%%%%%%%%% TITLE
\title{Self Supervised Low Dose Computed Tomography Image Denoising Using Invertible Network Exploiting Inter Slice Congruence}

\author{Sutanu Bera, Prabir Kumar Biswas\\
Indian Institute of Technology Kharagpur, India\\
\thispagestyle{empty}
{\tt\small sutanu.bera@iitkgp.ac.in, pkb@ece.iitkgp.ac.in}
% For a paper whose authors are all at the same institution,
% omit the following lines up until the closing ``}''.
% Additional authors and addresses can be added with ``\and'',
% just like the second author.
% To save space, use either the email address or home page, not both
}

\maketitle
%%%%%%%%% ABSTRACT
\begin{abstract}
The resurgence of deep neural networks has created an alternative pathway for low-dose computed tomography denoising by learning a nonlinear transformation function between low-dose CT (LDCT) and normal-dose CT (NDCT) image pairs. However, those paired LDCT and NDCT images are rarely available in the clinical environment, making deep neural network deployment infeasible. This study proposes a novel method for self-supervised low-dose CT denoising to alleviate the requirement of paired LDCT and NDCT images. Specifically, we have trained an invertible neural network to minimize the pixel-based mean square distance between a noisy slice and the average of its two immediate adjacent noisy slices. We have shown the aforementioned is similar to training a neural network to minimize the distance between clean NDCT and noisy LDCT image pairs. Again, during the reverse mapping of the invertible network, the output image is mapped to the original input image, similar to cycle consistency loss. Finally, the trained invertible network's forward mapping is used for denoising LDCT images. Extensive experiments on two publicly available datasets showed that our method performs favourably against other existing unsupervised methods.
\end{abstract}
\begin{figure}[t]
    \centering
    \includegraphics[width = 0.47\textwidth]{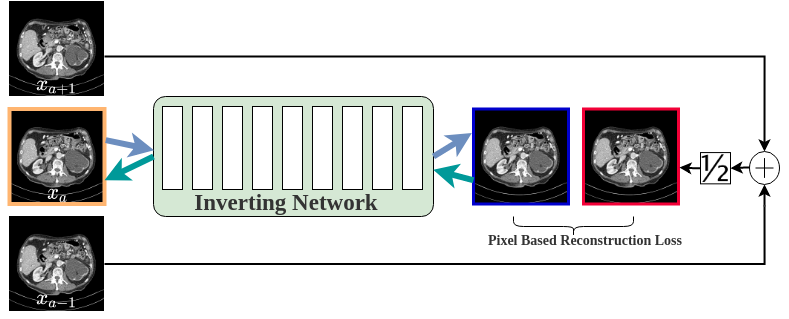}
    \caption{Illustration of our proposed method. During the forward pass the input image is mapped to the spatial average of two neighbouring slice, and during the reverse mapping the output image is mapped back to original noisy image. So the forward mapping is similar to Noise2Noise\cite{lehtinen2018noise2noise} paradigm, and the reverse mapping act as a regularizer enforcing no information loss.}
    \label{fig:abstract}
\end{figure}

%%%%%%%%% BODY TEXT
\section{Introduction}
 \thispagestyle{empty}
Low dose computed tomography (LDCT) imaging offers a nonvulnerable solution to the frequent requirement of computed tomography (CT) imaging by reducing the patient's radiation exposure at the time of imaging \cite{brenner2004radiation}, \cite{su2019low}. However, the lowering of radiation dose leads to degradation of the diagnostic quality of the low dose CT image, by creating several artifacts and noise. In the last few decades, researchers have tried to restore the diagnostic quality of LDCT images by various image post-processing methods. Among those methods, the deep neural network have been proven to be most efficient due to its commendable expressive power \cite{chen2017low}. Despite the superior image quality achieved by neural networks as compared to handcrafted prior functions, almost all the neural network-based methods require both noisy and clean images during the training. However, paired clean and noisy images are rarely available in practical scenarios. So, training deep networks without clean images is of crucial importance. Recently, there has been a surge in the studies exploring the possibilities of training deep networks without clean images for the purpose of denoising natural images corrupted with synthetic noise \cite{lehtinen2018noise2noise}, \cite{batson2019noise2self}, \cite{ulyanov2018deep}. However, we found these generalized solutions do not perform adequately in complicated signal-dependent CT noise or are ill-suited because of inconvenient training procedures.
Recently  Wu \textit{et al.} \cite{wu2019consensus} proposed a self supervised method for LDCT denoising based on the Noise2Noise \cite{lehtinen2018noise2noise} paradigm. However, in order to generate two independent realization of the same noisy image, their methods requires to divide the whole projection data into odd and even projection set. This heuristic approach lowers spatial resolution and signal strength of the input image, and also restricts the deployment of the technique due to interference with the workflow of existing CT scanners.
\par In this study, we have proposed a novel method for self supervised low dose CT denoising. Firstly, we have proposed to use the inter slice congruence for generating two noisy observation of the same image. We used the axial mean of the two opposite neighbouring slices to approximate the second noisy observation of the contemporary slice. However, the above mentioned method created a residual offset between two noisy observations. As a remedy we have leveraged the recently proposed concept of invertible neural network, which act both as a regularizer and a restoration network. Specifically, during the forward mapping we used the network to map a noisy axial slice to mean of its two opposite neighbouring slices, and during the reverse mapping the output is again mapped back to the original noisy image. The information lossless property of the invertible network work as a regularizer to overcome the residual offset which occurs due to axial approximation. 
\par We evaluated our proposed method on two datasets; namely, the 2016 NIH-AAPM-Mayo Clinic Low Dose CT Grand Challenge, and the ELCAP Public Lung Image Database. Extensive experiments on both these data sets demonstrated that our method is notably superior to other state-of-the-art unsupervised or self supervised low-dose CT denoising methods. 
\section{Comparison with other methods:} 
In recent times a lot of attention has been given to train neural network without clean images pairs. Among these methods, void networks or noise2self \cite{krull2019noise2void}, \cite{batson2019noise2self}, \cite{laine2019high} are the leaders for their performance in synthetic noise denoising, but we have shown these networks do not perform adequately in complicated CT noise. On the other hand, deep image prior \cite{ulyanov2018deep} is another self-supervised method to denoise image. However, the iterative deep image prior doesn't have any defined stopping criteria, which makes it unsuitable for medical applications; as with more iteration the network produces the original noisy image. The CycleGAN framework \cite{zhu2017unpaired} serves as the foundation for another well-known school of unsupervised denoising methods \cite{9059965}, \cite{li2020investigation}, \cite{huang2020cagan}.  Unpaired training data were utilized in these methods to train the denoising network. However, it is difficult and laborious to converge the networks trained with the CycleGAN framework, which makes these approaches inconvenient for widespread use. The chosen hyperparameter has an impact on denoising performance as well. Recently, D Wu \textit{et al.} \cite{wu2019consensus} proposed a consensus loss function for LDCT denoising; however, their methods require dividing the projection data into two sets before back-projection of sinogram data to reconstruct the input image. This restricts the deployment of the technique due to interference with the workflow of existing CT scanners. Whereas, our method is a simple plug-and-play method and does not interfere with workflow of the standard CT scanners.
\section{Method}
\subsection{Invertible Neural Network}
Invertible neural networks(INN) are originally designed for unsupervised learning of probabilistic models \cite{dinh2014nice}. These networks can transform a distribution to another distribution through a bijective function without losing information. Therefore, they are ideally suited to be used as normalizing flow \cite{kingma2018glow}. In recent work they have been used in various image to image translation task \cite{zhao2021invertible}, \cite{van2019reversible}, image denoising \cite{kwon2021cycle}, \cite{liu2021invertible}, image re-scaling \cite{xiao2020invertible}, image super resolution \cite{liang2021hierarchical}, etc.

INNs consist of layers that guarantee an invertible relationship between their input and output. Every invertible block, must satisfy the following condition, if $n = f_\theta(m)$, then $m = f^{-1}_\theta(n)$, where $f_\theta$ is the invertible block and $m$ and $n$ are the input and output to the block. In our study, we leveraged the invertible blocks formulation proposed in \cite{dinh2016density}. For the $l_{th}$ block the input $m$ is split into two groups $m_1$ and $m_2$ using channel wise splitting of same size. Then, they undergo an additive affine transformation with an identity branch augmentation as follows: 
\begin{equation}
    n_1 = m_1 + \phi_1(m_2) 
\end{equation} 
\begin{equation}
    n_2 = m_2 \odot \exp{\sigma(\phi_2(n_1))} + \phi_3(n_1)
\end{equation} 
Here, $\odot$ implies element wise multiplication. During the reverse mapping, given the output $n_1$ and $n_2$, the input $m_1$ and $m_2$ can computed as
\begin{equation}
    m_2 = (n_2 - \phi_3(n_1)) \odot \exp{-\sigma(\phi_2(n_1))} 
\end{equation} 
\begin{equation}
    m_1 = m_2 -  \phi_1(m_1)
\end{equation} 
Here, $\phi_1$, $\phi_2$, $\phi_3$ can be any function(e.g., convolutional layer) without any restriction on invertibility. The above formulation enable to directly determine inverse function $f^{-1}_\theta$ without any complicated computation or information loss.
\subsection{Inter Slice Congruence}
Let's consider $Y_1$, $Y_2$, $Y_3$ are three adjacent noisy slices, and $X_1$, $X_2$, $X_3$ are the corresponding clean CT slices. In general, $Y_i = X_i + \eta_i $, where $\eta_i$ is the random
variable of observed noise. We do have any pre-assumption on the characteristic of the observed noise. The ``$+$" sign signifies that there are observed noise at every pixel added with original signal. Our only assumption is $\eta_1$, $\eta_2$, and $\eta_3$, i.e., the observed noise in different axial slices are statistically independent from each other. Which is always true for CT images. \\
Ideally,
$X_1=X_2+\delta_1$.
Where $\delta_i$ is the offset which compensates for the small axial distance between two slices. Substituting the value of the $X_1$ we can get, $ Y_1 - \eta_1 = X_2 + \delta_1$, or $Y_1 = X_2 + \delta_1 + \eta_1$. 
Similarly from $X_2=X_3+\delta_2$, we can get $Y_3 = X_2 - \delta_2 +\eta_3$. 
Adding the above two expressions, we get
%\[Y_1 + Y_3  = 2*X_2+ (\delta_1 - \delta_2) + (\eta_1 + \eta_3)\]
\begin{equation}
 \frac{1}{2} (Y_1+Y_3) = X_2 + \frac{1}{2}(\delta_1 - \delta_2) + \frac{1}{2}(\eta_1 + \eta_3)  
\end{equation}
or,
\begin{equation}
    \frac{1}{2} (Y_1+Y_3) \approx X_2 + \eta_2^\prime  
\end{equation}
Here, $\eta_2^\prime = \frac{1}{2}(\eta_1 + \eta_3)$. The value of $\frac{1}{2}(\delta_1 - \delta_2)$ is very small as compared to the $X_2$, so we can ignore this term (we will discuss more about it in section 2.3).  A deep neural network $f(\hspace{5pt};\theta)$, parameterized by $\theta$, can be trained to minimize the following distance function:
\begin{equation}
    \theta^* = \operatorname*{argmin}_\theta \Big \{ d\big\{f(Y_2; \theta) , \frac{1}{2}(Y_1+Y_3)\big\} \Big\}
\end{equation}
or,
\begin{equation}
    \theta^* = \operatorname*{argmin}_\theta \Big \{ d\big\{f(Y_2; \theta) , X_2 + \eta_2^\prime \big\} \Big\}
\end{equation}
or,
\begin{equation}
    \theta^* = \operatorname*{argmin}_\theta \Big \{ d\big\{f(X_2 + \eta_2; \theta) , X_2 + \eta_2^\prime \big\} \Big \}
\end{equation}	
According to Lehtinen \etal \cite{lehtinen2018noise2noise}, training a neural network to minimize the distance between two independent noisy realizations of the same clean signal is equivalent to training a neural network by minimizing the distance between the original clean signal and noisy signal. 
So, 
    \begin{equation}
        \theta^* = \operatorname*{argmin}_\theta \Big \{ d\big\{f(Y_2; \theta) , X_2 \} \Big \}
    \end{equation} 
    \begin{table*}[]
\centering
\resizebox{0.9\textwidth}{!}{%
\begin{tabular}{|l|l|}
\hline
Layer Name & Details \\ \hline
$\mathcal{E}_Y$ & Single Covolutional Layer with 1 input channel and 64 output channel. Kernel size $3 \times 3$. \\ \hline
$\mathcal{E}_X$ & Single Covolutional Layer with 1 input channel and 64 output channel. Kernel size $3 \times 3$. \\ \hline
$\mathcal{D}_X$ & Single Covolutional Layer with 64 input channel and 1 output channel. Kernel size $3 \times 3$. \\ \hline
$\mathcal{D}_Y$ & Single Covolutional Layer with 64 input channel and 1 output channel. Kernel size $3 \times 3$. \\ \hline
$\phi_s$ & Dense Block \cite{huang2017densely} with grow rate of 32, and four layer. Kernel of size $3 \times 3$, 64 input and output channel.\\ \hline
\end{tabular}%
}
\caption{Details of every component used in our method. }
\label{tab:arch}
\end{table*}
\begin{figure*}[ht]
    \centering
    \includegraphics[width=\textwidth]{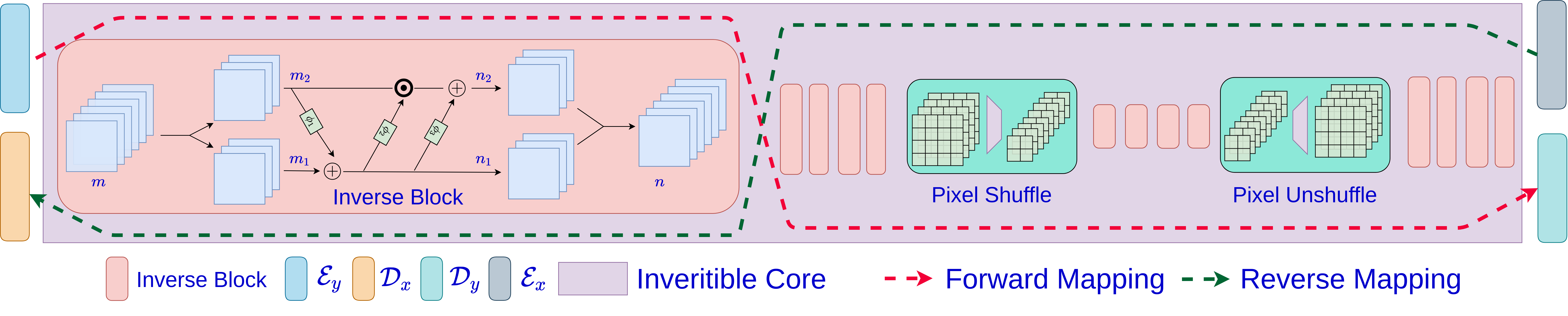}
    \caption{Simplified Illustration of Proposed Network. A detailed layer-wise description of the network is given in Table \ref{tab:arch}}
    \label{fig:inv}
\end{figure*}
\subsection{Design Concept}
We have given a graphical representation of our proposed method in Figure \ref{fig:abstract}. As shown in the Figure, the noisy input image is mapped to the average of its two neighboring slices during the forward mapping. As discussed in the above section, it is similar to training a neural network to map a noisy image to the noise-free image counterpart. Although we said the offset $\delta_1 - \delta_2$ is negligible, but in some cases, it may become considerable, especially at the beginning and the end of the organ (we have shown an example in the section 4.1). To regularize the network in this end organ case, we again mapped the output image back to the noisy input image during the reverse mapping of the invertible network, similar to cycle consistency loss. Furthermore, the architectures of the invertible network by design satisfy cycle-consistency before training even begins, which again regulates the network from the interslice smoothing at the end of organ cases. In our design, we employed two encoders ($\mathcal{E}_X$,$\mathcal{E}_Y$), two decoders ($\mathcal{D}_X$,$\mathcal{D}_Y$), and a invertible core $\mathcal{I}$. The encoder $\mathcal{E}_Y$ takes the input noisy slice $Y_i$ as the input and produces a higher dimension overcomplete 3D feature map $\Tilde{Y_i} \in \mathbb{R}^{C \times W \times H}$, i.e., $\Tilde{Y_i} = \mathcal{E}_Y(Y_i)$. Here, $W$ and $H$ are the width and the height of the input noisy image patch, and $C$ is the number of channels in the feature map. This feature map $\Tilde{Y_i}$ is used as the input to the invertible core $\mathcal{I}$ during the forward map. The invertible core $\mathcal{I}$ produces an output $\tilde{\tilde{Y_i}} \in \mathbb{R}^{C \times W \times H}$. The decoder $\mathcal{D}_Y$ projects the higher dimensional feature map $\tilde{\tilde{Y_i}}$  to image space. Thus, the output of the decoder $\mathcal{D}_Y$ is the predicted clean image $\hat{X_i}$, i.e., $\hat{X_i} = \mathcal{D}_Y(\tilde{\tilde{Y_i}})$. During the reverse mapping, the encoder $\mathcal{E}_X$ first takes the predicted clean image $\hat{X_i}$ as the input and produces a higher dimension overcomplete 3D feature map $\Tilde{X_i} \in \mathbb{R}^{C \times W \times H}$, i.e., $\Tilde{X_i} = \mathcal{E}_X(\hat{X_i})$. Then, this feature map is transformed into a higher dimensional feature map $\tilde{\tilde{X_i}} \in \mathbb{R}^{C \times W \times H}$ using the reverse invertible core $\mathcal{I}^{-1}$. Finally the decoder $\mathcal{D}_X$ project $\tilde{\tilde{X_i}}$ into predicted noisy image $\hat{Y_i}$. So the mapping for the forward and reverse processes are given by
\[ Forward = \mathcal{D}_Y \circ \mathcal{I} \circ  \mathcal{E}_Y  \]
\[ Reverse = \mathcal{D}_X \circ \mathcal{I}^{-1} \circ \mathcal{E}_X \]
The invertible core $\mathcal{I}$ and its inverse $\mathcal{I}^{-1}$ shares parameters with each other. Essentially training the $\mathcal{I}^{-1}$ will automatically train $\mathcal{I}$ and vise versa. Via training the invertible network for reverse mapping we are regulating the information loss which may happen in the forward mapping in the end organ cases, as for the reverse mapping we have exact pixel by pixel correspondence between the target noisy image and predicted noisy image. 
\\A visual illustration of the proposed network is given in Figure \ref{fig:inv}. We used pixel shuffle and unshuffle layers to downsample and upsample the feature maps, as these two layers satisfy the invertibility property. We used total $12$ invertibles blocks in total. A detailed layer-wise description of the network is given in Table \ref{tab:arch}.

\subsection{Learning}
We used mean square error as the reconstruction loss in our study. In the forward mapping the loss is calculated between the predicted clean image $\hat{X_i}$ and average of two neighbouring noisy slice, i.e.
\begin{equation}
    \mathcal{L}_f= \frac{1}{k} \Big\|\frac{1}{2}(Y_{i-1}+Y_{i+1}) - \hat{X_i}\Big\|^2_2
\end{equation}
or,
\begin{equation}
    \mathcal{L}_f= \frac{1}{k} \Big\|\frac{1}{2}(Y_{i-1}+Y_{i+1}) - \mathcal{E}_Y \circ \mathcal{I} \circ \mathcal{D}_Y (Y_i)\Big\|^2_2
\end{equation}
During reverse mapping the loss is calculated between the predicted noisy image $\hat{Y_i}$ and original noisy image $Y_i$, i.e.
\begin{equation}
    \mathcal{L}_r= \frac{1}{k} \Big\|Y_i - \hat{Y_i}\|^2_2
\end{equation}
or,
\begin{equation}
    \mathcal{L}_r= \frac{1}{k} \Big\|Y_i - \mathcal{E}_X \circ \mathcal{I}^{-1} \circ \mathcal{D}_X(\hat{X_i})\Big\|^2_2
\end{equation}
or,
\begin{equation}
    \mathcal{L}_r= \frac{1}{k} \Big\|Y_i - \mathcal{E}_X \circ \mathcal{I}^{-1} \circ \mathcal{D}_X\Big(\mathcal{E}_Y \circ \mathcal{I} \circ \mathcal{D}_Y (Y_i)\Big)\Big\|^2_2
\end{equation}
Here $k$ is the total number of samples in every batch. The total loss $\mathcal{L} = \mathcal{L}_f + \mathcal{L}_r$ is back-propagated for each batch to train the network. Note that the loss $\mathcal{L}_f$ never minimizes completely during the training, as here the network is asked to solve an impossible task, i.e., transform a noisy version of an image into another noisy version. After training, the network produces images with the smallest average deviation from the noisy target images, and identical to the general class of deviation-minimizing estimator \cite{huber1992robust}, \cite{lehtinen2018noise2noise}; this estimated image turns out to be the original clean image.

\begin{figure*}
    \begin{subfigure}[]{0.19\textwidth}
        \centering
        \includegraphics[width=\textwidth]{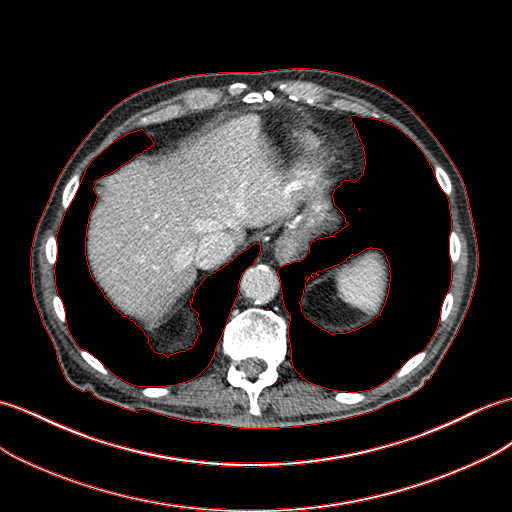}
        \caption{LDCT}
        \label{fig:y equals x}
    \end{subfigure}
    \begin{subfigure}[]{0.19\textwidth}
        \centering
        \includegraphics[width=\textwidth]{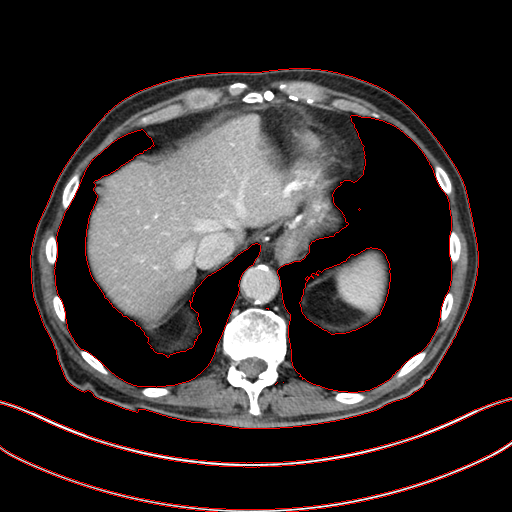}
        \caption{NDCT}
        \label{fig:three sin x}
    \end{subfigure}
    \begin{subfigure}[]{0.19\textwidth}
        \centering
        \includegraphics[width=\textwidth]{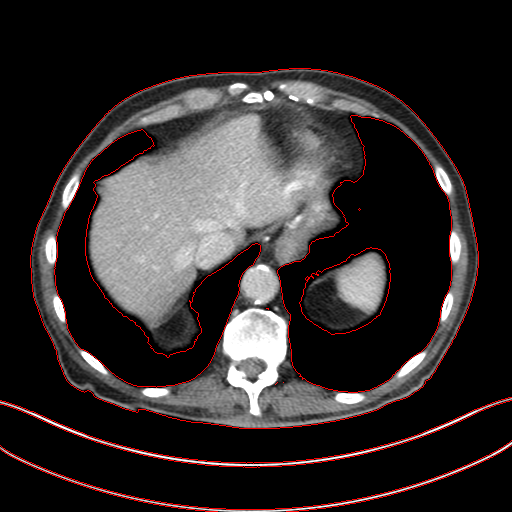}
        \caption{M1}
    \end{subfigure}
    \begin{subfigure}[]{0.19\textwidth}
        \centering
        \includegraphics[width=\textwidth]{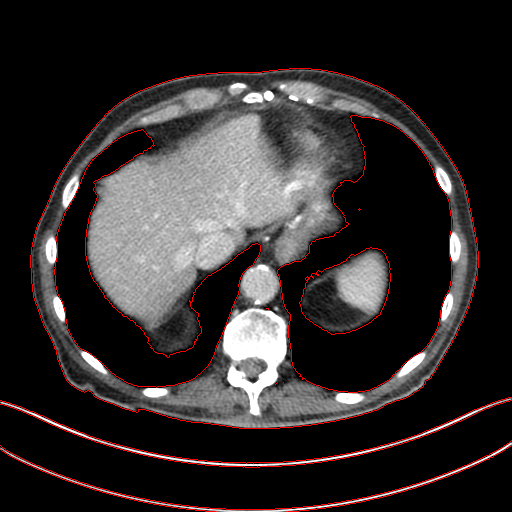}
        \caption{M2}
    \end{subfigure}
    \begin{subfigure}[]{0.19\textwidth}
        \centering
        \includegraphics[width=\textwidth]{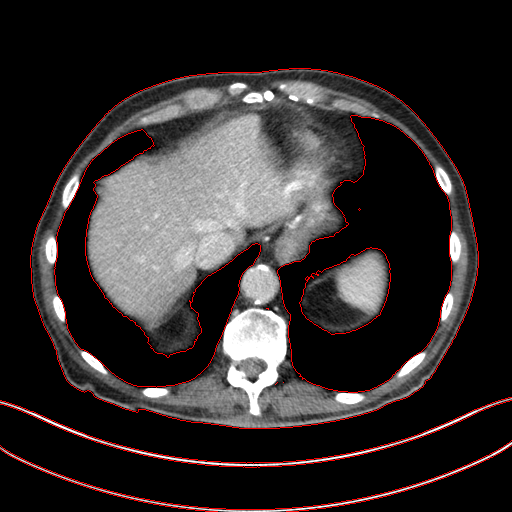}
        \caption{M3}
    \end{subfigure}
    \caption{Comparison of denoising performance of different network. The boundary line of different organ extracted from the NDCT image is superimposed on other images. As seen, many pixels are missing around the boundary line in the output of M1}
    \label{fig:abla}
\end{figure*}

\begin{figure*}
\centering
    \begin{subfigure}[]{0.17\textwidth}
        \centering
        \includegraphics[width=\textwidth]{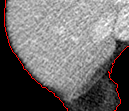}
        \caption{LDCT}
        \label{fig:y equals x}
    \end{subfigure}
    \begin{subfigure}[]{0.17\textwidth}
        \centering
        \includegraphics[width=\textwidth]{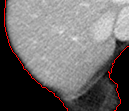}
        \caption{NDCT}
        \label{fig:three sin x}
    \end{subfigure}
    \begin{subfigure}[]{0.17\textwidth}
        \centering
        \includegraphics[width=\textwidth]{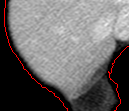}
        \caption{M1}
    \end{subfigure}
    \begin{subfigure}[]{0.17\textwidth}
        \centering
        \includegraphics[width=\textwidth]{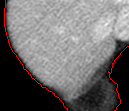}
        \caption{M2}
    \end{subfigure}
    \begin{subfigure}[]{0.17\textwidth}
        \centering
        \includegraphics[width=\textwidth]{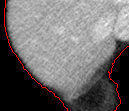}
        \caption{M3}
    \end{subfigure}
    \caption{Zoomed version of the ROI taken from the images of Figure \ref{fig:abla}. In the zoomed version missing boundary pixel is clearly visible in the output of M1.}
    \label{fig:abla3}
\end{figure*}

\begin{figure*}
    \centering
    \begin{subfigure}[]{0.19\textwidth}
        \centering
        \includegraphics[width=\textwidth]{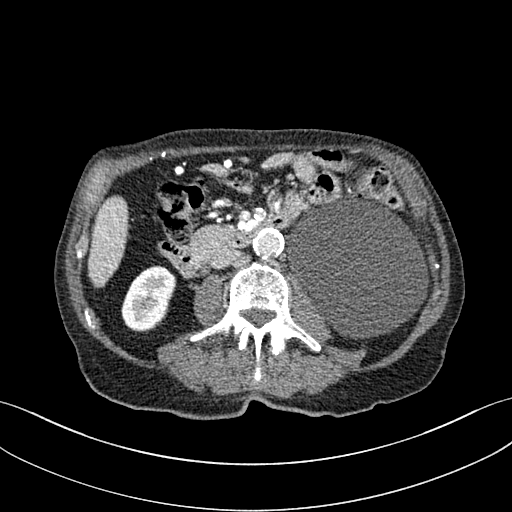}
        \caption{Original LDCT}
        \label{fig:y equals x}
    \end{subfigure}
    \begin{subfigure}[]{0.19\textwidth}
        \centering
        \includegraphics[width=\textwidth]{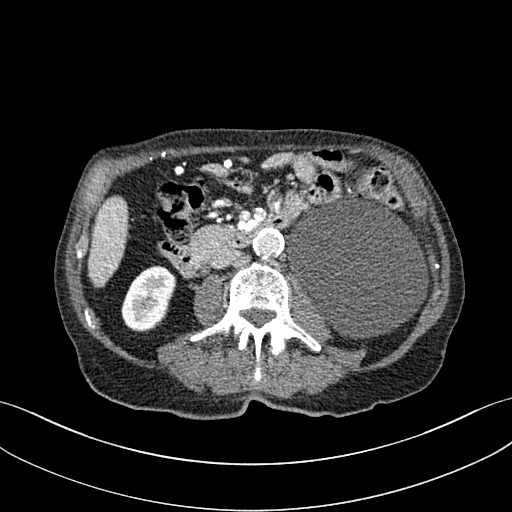}
        \caption{Predicted LDCT}
        \label{fig:three sin x}
    \end{subfigure}
    \begin{subfigure}[]{0.19\textwidth}
        \centering
        \includegraphics[width=\textwidth]{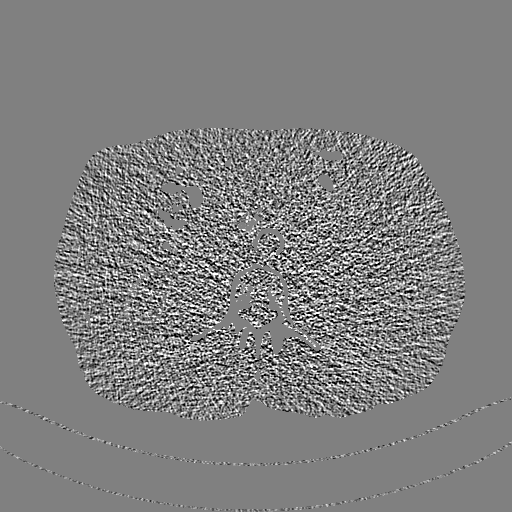}
        \caption{Original Noise }
    \end{subfigure}
    \begin{subfigure}[]{0.19\textwidth}
        \centering
        \includegraphics[width=\textwidth]{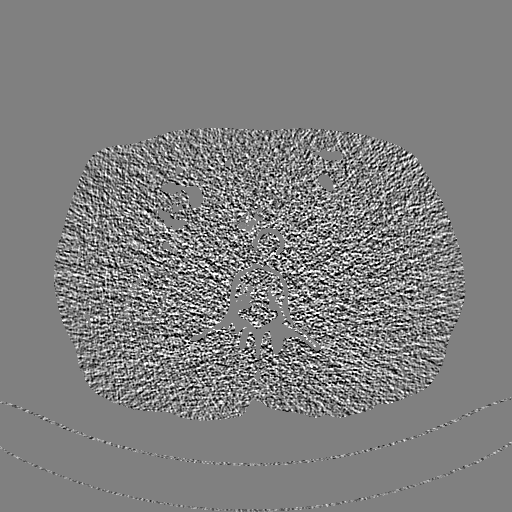}
        \caption{Predicted Noise }
    \end{subfigure}
    \caption{Performance of invertible network in reverse mapping. The display window for CT images is [$ - 160$, 240] HU, and for noise pattern [$ - 50$, 50] HU}
    \label{fig:abla2}
\end{figure*}

\section{Experimental Settings}
We evaluated the proposed method on one synthesized dataset and one clinical datasetst referred as D1, and D2, described below:
\begin{enumerate*}
    \item \textbf{D1:} 2016 NIH-AAPM-Mayo Clinic Low Dose CT Grand Challenge Dataset
    \item \textbf{D2:} ELCAP Public Lung Image Database.
\end{enumerate*}
The first dataset contains simulated quarter dose CT scan of 10 patients. For each patient, there are approximately 250 slices per scan. The second data set includes real low-dose CT scans from 50 patients.
We used Adam optimizer with a batch size of 16. The learning rate was initially set to $1e^{-4}$ and was set to decrease by a factor of 2 after every 6000 iterations. For training, we have taken ten randomly cropped patches from each slice, each patch of size $120 \times 120$. The benchmark data set D1 contains paired NDCT and LDCT images, so we chose this dataset for the objective evaluation of our proposed method, and D2 is used for subjective evaluation only. For dataset D1, we used CT slices from seven patients as the training set and used the remaining three patients' data as the test set. For dataset D2, we used images from 30 patients as our training data and followed the same training procedure as mentioned above.
\section{Result and Discussion}
\subsection{Ablation Study}
This section systemically investigates the efficacy of every module proposed in this study. We considered three different networks; first, baseline model(M1), where the inverting block is replaced with dense block, and trained using minimizing mean square distance between two noisy observations(i.e. $\mathcal{L}_f$). Next, two independent baseline model(one for forward mapping, one for reverse mapping) is jointly trained using a linear combination $\mathcal{L}_f$, and $\mathcal{L}_r$, similar to cycle consistent network paradigm. The forward mapping network is used for testing. We refer this model as M2. In both M1, and M2, we increased the depth of the network to make the representation power of these networks comparable with inverting network. Finally, the proposed method, referred as M3. Table \ref{tab:abla} depicts the objective evaluation of the three networks using the D1 dataset. Both M2 and M3 use reverse mapping to regularize the network; the influence of the same in the denoising performance is evident from Table \ref{tab:abla}. Adding cycle consistency loss has improved the performance of the same baseline model significantly. The inverting network performed considerably better than the network M2. It improves PSNR by 0.23dB. As discussed in the above section, in case cycle consistency loss, an additional network is trained, but that does not always guarantee invertibility, whereas inverting network architecture inherently possesses reversibility, which acts as a strong regularizer. In Figure \ref{fig:abla} we have shown denoising performance of different networks visually. To demonstrate the requirement of regularization, we first extract the boundary line of the various organ from the NDCT image and superimpose the boundary line on the output of different networks. As shown in Figure \ref{fig:abla}, many pixels around the boundary line of M1 network output are missing. Using the reverse mapping, the issue of the end-organ missing pixel is successfully overcome in M2 and M3. The granular pattern is also less present in the M3 than M2. The zoomed version of a ROI taken from the images of Figure \ref{fig:abla} is given in Figure \ref{fig:abla3} for better perception. In Figure \ref{fig:abla2} we give an example of the performance of the invertible network in reverse mapping. Here, the predicted LDCT image is produced by using the predicted clean image of the forward mapping as the input for reverse mapping. As shown, the predicted noisy pattern is similar to the original noise pattern. The same streaking artifacts are present in both the noise pattern; also, the noise variance is different in the various spatial region depending on the signal intensity of the original CT image. It validates that the loss of information in reverse mapping is minimal. Due to the invertible network's structural advantage, the network also preserves every information present in the input image in the forward mapping.

%compare the proposed method with five state-of-the-art methods that are closely related to our problem. The first two are the generalized solution for training deep neural network without clean images: \begin{enumerate*}
%    \item Noise2void\cite{laine2019high}, \item Deep Image Prior\cite{ulyanov2018deep}.
%\end{enumerate*} Next, we considered the current state-of-the-art method for the self-supervised LDCT denoising. We will refer to it as ConsensusNet\cite{wu2019consensus}. Next, BM3D\cite{dabov2006image} and finally a state-of-the-art supervised learning method for CT denoising, \ie,WGAN\cite{yang2018low}.
\subsection{Performance on the simulated data}
\begin{figure*}
    \centering
    \begin{subfigure}[]{0.21\textwidth}
        \centering
        %includegraphics[origin=c]{file}
        \includegraphics[width=\textwidth]{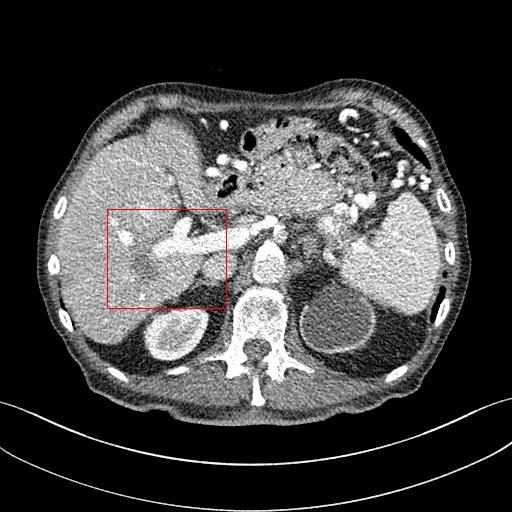}
        \caption{LDCT}
        \label{fig:y equals x}
    \end{subfigure}
    \begin{subfigure}[]{0.21\textwidth}
        \centering
        \includegraphics[width=\textwidth]{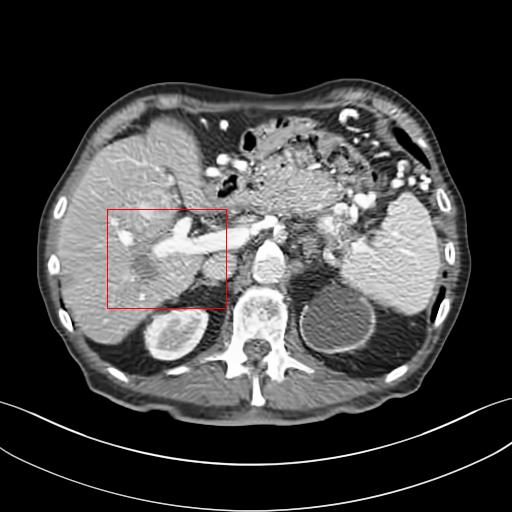}
        \caption{BM3D}
        \label{fig:three sin x}
    \end{subfigure}
    \begin{subfigure}[]{0.21\textwidth}
        \centering
        \includegraphics[width=\textwidth]{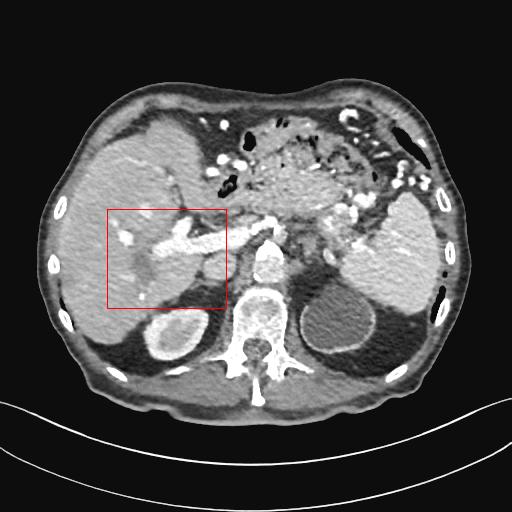}
        \caption{DIP}
    \end{subfigure}
    \begin{subfigure}[]{0.21\textwidth}
        \centering
        \includegraphics[width=\textwidth]{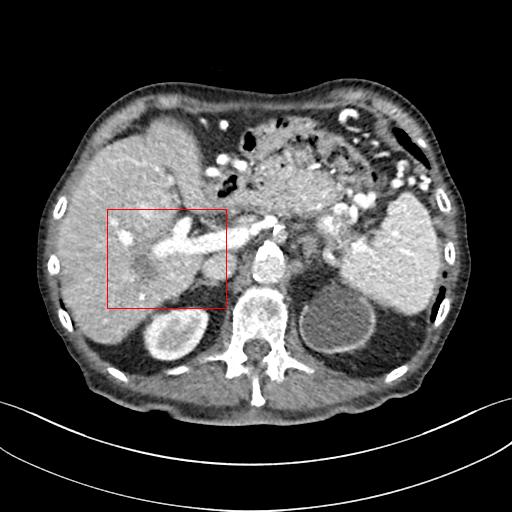}
        \caption{CycleGAN}
    \end{subfigure}
    \begin{subfigure}[]{0.21\textwidth}
        \centering
        \includegraphics[width=\textwidth]{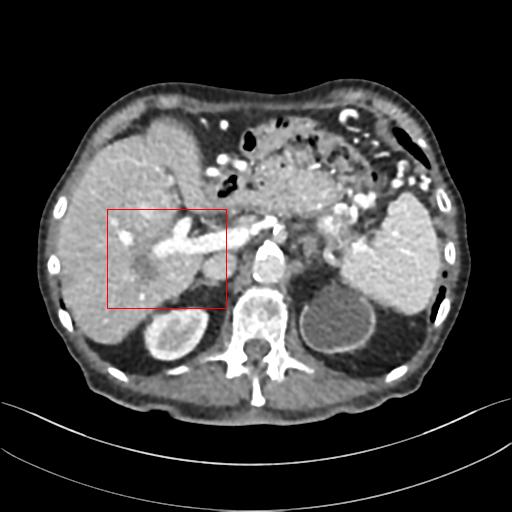}
        \caption{ConsensusNet}
    \end{subfigure}
    \begin{subfigure}[]{0.21\textwidth}
        \centering
        \includegraphics[width=\textwidth]{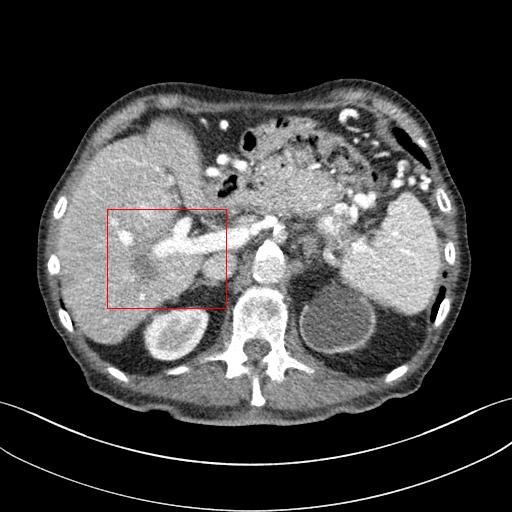}
        \caption{Proposed}
    \end{subfigure}
    \begin{subfigure}[]{0.21\textwidth}
        \centering
        \includegraphics[width=\textwidth]{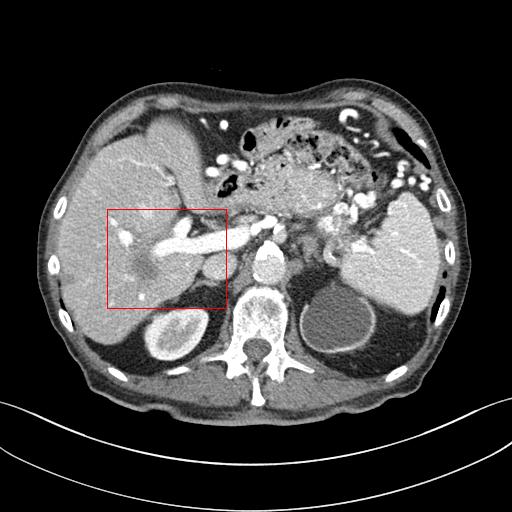}
        \caption{NDCT}
    \end{subfigure}
    \caption{Result of denoising for comparison. We have shown an example of denoising performance on image taken from the 2016 NIH-AAPM-Mayo Clinic Grand Challenge dataset. The display window is [$ - 140$, 260] HU for better visualization of low attenuated lesion. Readers are requested to zoom in for better view.}
    \label{fig:mayo_full}
\end{figure*}

\begin{figure*}
    \centering
    \begin{subfigure}[]{0.21\textwidth}
        \centering
        %includegraphics[origin=c]{file}
        \includegraphics[width=\textwidth]{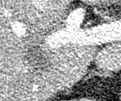}
        \caption{LDCT}
        \label{fig:y equals x}
    \end{subfigure}
    \begin{subfigure}[]{0.21\textwidth}
        \centering
        \includegraphics[width=\textwidth]{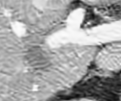}
        \caption{BM3D}
        \label{fig:three sin x}
    \end{subfigure}
    \begin{subfigure}[]{0.21\textwidth}
        \centering
        \includegraphics[width=\textwidth]{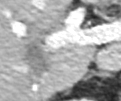}
        \caption{DIP}
    \end{subfigure}
    \begin{subfigure}[]{0.21\textwidth}
        \centering
        \includegraphics[width=\textwidth]{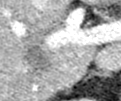}
        \caption{CycleGAN}
    \end{subfigure}
    \begin{subfigure}[]{0.21\textwidth}
        \centering
        \includegraphics[width=\textwidth]{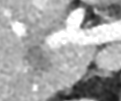}
        \caption{ConsensusNet}
    \end{subfigure}
    \begin{subfigure}[]{0.21\textwidth}
        \centering
        \includegraphics[width=\textwidth]{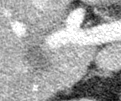}
        \caption{Proposed}
    \end{subfigure}
    \begin{subfigure}[]{0.21\textwidth}
        \centering
        \includegraphics[width=\textwidth]{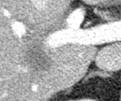}
        \caption{NDCT}
    \end{subfigure}
    \caption{Comparison of denoising performance of different network. Zoomed view of the region inside the bounding box shown in the images for Figure \ref{fig:mayo_full}}
    \label{fig:mayo_crop}
\end{figure*}

We compare performance of our method against four state of the art unsupervised method; namely, DIP \cite{ulyanov2018deep}, ConsensusNet \cite{wu2019consensus}, CycleGAN \cite{9059965}, BM3D \cite{dabov2006image}. Performance metrics used are, peak signal to noise ratio (PSNR), structural similarity index measure (SSIM), fréchet inception distance (FID) \cite{heusel2017gans}, perceptual loss (PL) \cite{johnson2016perceptual}, texture matching loss (TML) \cite{sajjadi2017enhancenet}, and visual information fidelity (VIF) \cite{xuan2004reversible}. 
\begin{table}[h]
\centering
\resizebox{0.2\textwidth}{!}{%
\begin{tabular}{|l|l|l|}
\hline
Method & PSNR  & SSIM  \\ \hline
M1     & 30.80 & 0.891 \\ \hline
M2     & 31.03 & 0.899 \\ \hline
M3     & 31.26 & 0.908 \\ \hline
\end{tabular}%
}
\caption{Comparison of different trained networks in terms of PSNR and SSIM.}
\label{tab:abla}
\end{table}
\begin{table*}[]
\centering
\resizebox{0.75\textwidth}{!}{%
\begin{tabular}{|l|c|c|c|c|c|c|}
\hline
Method      & PSNR $\uparrow$ & SSIM $\uparrow$ & FID $\uparrow$  & TML $\uparrow$  & PL $\uparrow$ & VIF  $\uparrow$\\ \hline
DIP  & 28.08 $(\pm 2.82)$ & 0.848 $(\pm 0.07)$ & 58.96 $(\pm 3.12)$ &  0.402 $(\pm 0.012)$    & 0.00189 $(\pm 2.3e-5)$  & 0.652 $(\pm 0.10)$  \\ \hline
ConsensusNet & 29.02 $(\pm 1.82)$ & 0.881 $(\pm 0.05)$ & 39.47 $(\pm 1.89)$ &  0.079  $(\pm 0.007)$    & 0.00171 $(\pm 4.7e-5)$   & 0.711 $(\pm 0.06)$  \\\hline
BM3D & 30.02 $(\pm 1.95)$ & 0.885 $(\pm 0.06)$ & 49.23 $(\pm 2.75)$ & 0.375 $(\pm 0.012)$ & 0.00152 $(\pm 2.9e-5)$ &  0.732 $(\pm 0.05)$  \\ \hline
CycleGAN        & 30.86 $(\pm 1.69)$ & 0.886 $(\pm 0.03)$ & 30.85 $(\pm 1.31)$ & 0.038 $(\pm 0.005)$ &  \textbf{0.00054} $(\pm 0.2e-5)$  & 0.955 $(\pm 0.07)$ \\ \hline
Proposed    &   \textbf{31.26} $(\pm 1.52)$    &    \textbf{0.893} $(\pm 0.03)$  & \textbf{30.23} $(\pm 0.89)$ &  \textbf{0.012} $(\pm 0.003)$     &  0.00068 $(\pm 0.3e-5)$  & \textbf{0.989} $(\pm 0.05)$ \\ \hline
\end{tabular}%
}

\caption{Comparison with other methods in terms of PSNR, SSIM, FID, TML, PL, and VIF. The results are obtained by taking the average of all the images of the test set. $\uparrow$ indicates higher values are better. Best results are marked in bold.}
\label{tab:result}
\end{table*}
As shown in Table \ref{tab:result}, DIP has exhibited the worst performance among all the methods. On the other hand, the ConsensusNet yielded a better FID and TML than BM3D but a lower PSNR and SSIM than other methods. The ConsensusNet divided the original projection data of the low dose CT image into two subgroups and back-projected to create the noisy input signal. Consequently, the noisy input image is much noisier than the original LDCT image. Also, the structural loss occurred during the generation of noisy images, so as a result, the PSNR and SSIM of this method are lower than other methods. Texture matching loss (TML) is used to measure the texture difference between the reconstructed and original images. The lower value of TML indicates that the generated texture is similar to the original. In comparison, FID estimates the distance between the distribution of the generated image and real images. A lower value of FID signifies the generated images are more similar to the original image. The current deep learning era demands a denoised image with a low value of these metrics. These denoised images may be used as input for other image classification tasks or segmentation networks. In this regard, the ConsensusNet is superior to the BM3D because it uses the deep neural network's expression power. CycleGAN is another powerful unsupervised method for image-to-image translation; it achieved better performance than the other methods. However, CycleGAN has a lot of bottlenecks, e.g., longer training time, computation power, hyper-parameter tuning, etc. All these bottlenecks make CycleGAN ill-suited for practical deployment. Meanwhile, our proposed method has achieved the highest PSNR, SSIM, FID, and VIF among all the other methods.
Next, we compare the result of denoising visually in Figure \ref{fig:mayo_full}. It can be observed that the proposed method performs significantly better than the other unsupervised methods. BM3D output produced a blurry denoised image and contained many splotchy artifacts. The same blurriness can be observed in the output of ConsensusNet, and DIP, although noise suppression is adequate, and splotchy artifact is absent. In the output of CycleGAN, we can observe the presence of residual noise, especially in the high noise regions. Next, we identified one low attenuated lesion in the sample image and marked the lesion with a red colour bounding box. The zoomed view of the region inside the bounding box is given in Figure \ref{fig:mayo_crop}. In our method's output image, the lesion's visibility is enhanced significantly than in other methods. Despite being an unsupervised method, the visibility of the lesion is comparable with the original NDCT image. Also, from the zoomed view, we can perceive that our method has suppressed the granular pattern without losing the original image's texture.

\begin{comment}
\begin{figure*}[h]
    \centering
    \includegraphics[width=0.95\textwidth]{self_result.drawio.png}
    \caption{Results of denoising for comparison. (Clockwise from top-left: LDCT, BM3D, Deep Image Prior, CycleGAN, NDCT, Proposed, ConsensusNet). We have identified a low attenuated lesion in the slice, and gave the zoomed region in the insect. The display window is [$ - 160$, 240] HU.}
    \label{fig:mayo}
\end{figure*}

\begin{figure*}
    \centering
    \includegraphics[width=.95\textwidth]{real_icip.drawio.png}
    \caption{Result of denoising for comparison. (Clockwise from top-left: LDCT, BM3D, Deep Image Prior, CycleGAN, Proposed, ConsensusNet.) The display window is [$ - 175$, 260] HU.}
    \label{fig:real_result}
\end{figure*}
\end{comment}

\begin{figure*}
    \centering
    \begin{subfigure}[]{0.21\textwidth}
        \centering
        \includegraphics[width=\textwidth]{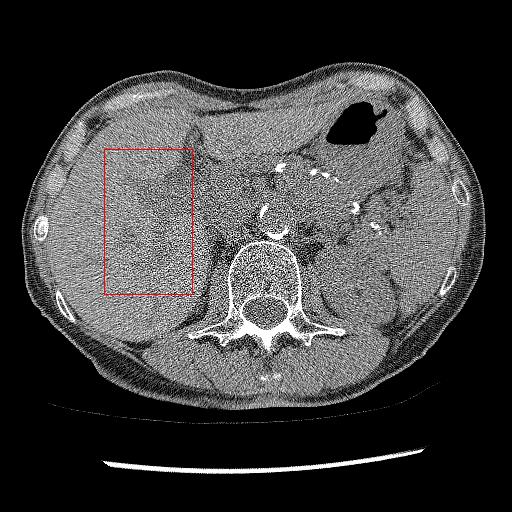}
        \caption{LDCT}
        \label{fig:y equals x}
    \end{subfigure}
    \begin{subfigure}[]{0.21\textwidth}
        \centering
        \includegraphics[width=\textwidth]{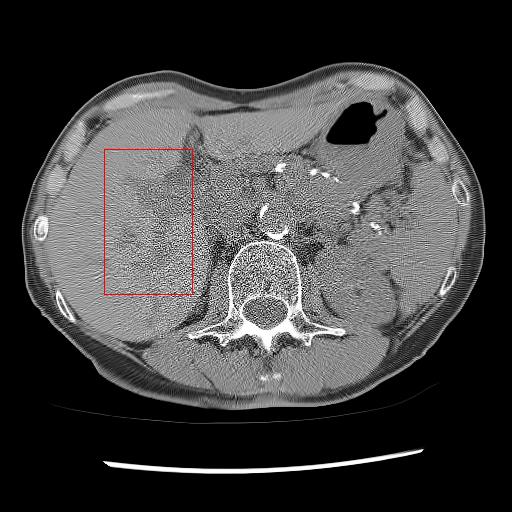}
        \caption{BM3D}
        \label{fig:three sin x}
    \end{subfigure}
    \begin{subfigure}[]{0.21\textwidth}
        \centering
        \includegraphics[width=\textwidth]{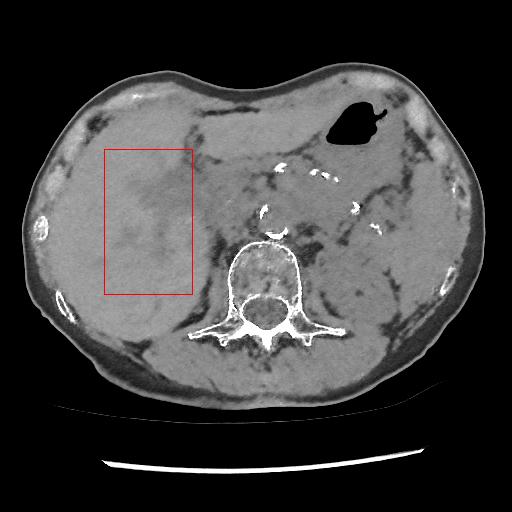}
        \caption{DIP}
    \end{subfigure}
    \begin{subfigure}[]{0.21\textwidth}
        \centering
        \includegraphics[width=\textwidth]{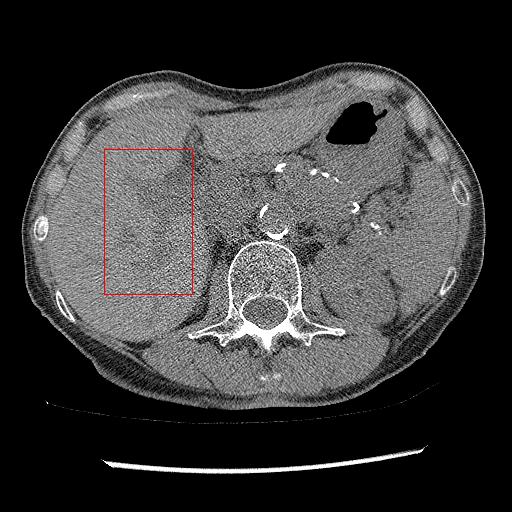}
        \caption{CycleGAN}
    \end{subfigure}
    \begin{subfigure}[]{0.21\textwidth}
        \centering
        \includegraphics[width=\textwidth]{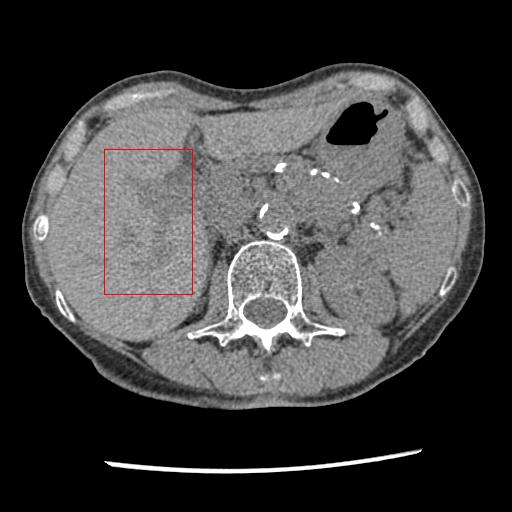}
        \caption{ConsensusNet}
    \end{subfigure}
        \begin{subfigure}[]{0.21\textwidth}
        \centering
        \includegraphics[width=\textwidth]{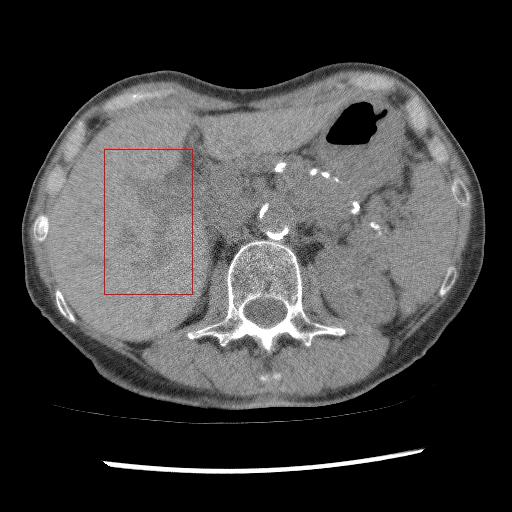}
        \caption{Proposed}
    \end{subfigure}
    \caption{Result of denoising for comparison. We have shown an example of denoising performance on image taken from the ELCAP Public Lung Image
    Database. The display window is [$ - 175$, 240] HU. Readers are requested to zoom in for better view.}
    \label{fig:real_full}
\end{figure*}

\begin{figure*}
    \centering
    \begin{subfigure}[]{0.201\textwidth}
        \centering
        %includegraphics[origin=c]{file}
        \includegraphics[angle=90,width=\textwidth]{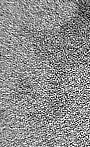}
        \caption{LDCT}
        \label{fig:y equals x}
    \end{subfigure}
    \begin{subfigure}[]{0.201\textwidth}
        \centering
        \includegraphics[angle=90,width=\textwidth]{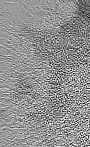}
        \caption{BM3D}
        \label{fig:three sin x}
    \end{subfigure}
    \begin{subfigure}[]{0.201\textwidth}
        \centering
        \includegraphics[angle=90,width=\textwidth]{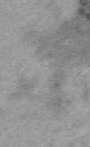}
        \caption{DIP}
    \end{subfigure}
    \begin{subfigure}[]{0.201\textwidth}
        \centering
        \includegraphics[angle=90,width=\textwidth]{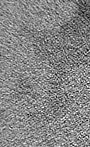}
        \caption{CycleGAN}
    \end{subfigure}
    \begin{subfigure}[]{0.201\textwidth}
        \centering
        \includegraphics[angle=90,width=\textwidth]{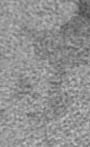}
        \caption{ConsensusNet}
    \end{subfigure}
        \begin{subfigure}[]{0.201\textwidth}
        \centering
        \includegraphics[angle=90,width=\textwidth]{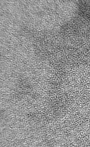}
        \caption{Proposed}
    \end{subfigure}
    \caption{Comparison of denoising performance of different network. Zoomed view of the region inside the bounding box shown in the images for Figure \ref{fig:real_full}}
    \label{fig:real_crop}
\end{figure*}

\begin{comment}
\begin{table}[]
\resizebox{0.47\textwidth}{!}{%
\begin{tabular}{ccccccc}\hline
 & LDCT & BM3D & DIP & ConsensusNet & CycleGAN & Proposed \\\hline
ROI1(RED) & 0.84 & 0.65 & 0.49 & 0.54 & 0.70 & \textbf{0.42} \\
ROI2(GREEN) & 0.87 & 0.70 & 0.54 & 0.61 & 0.76 & \textbf{0.49} \\\hline
\end{tabular}%
}
\caption{Comparison between different method in real low dose ct image. The variance of the pixel of two marked region in the Figure \ref{fig:real_result} for different method is reported.}
\label{tab:real}
\end{table}
\end{comment}

\subsection{Performance on the real low dose CT data}
Here, we analyse the performance of the proposed method on clinical data. Since there are no ground truths available, only qualitative comparisons are performed.  
%First, we consider a practical situation where only low-dose CT data from one patient is available. We can train and test the model on the same data for denoising the LDCT images under this condition. We trained our network on three random patients from the ELCAP Public Lung Image Database(with id $W0001, W0002, W0019$) for this analysis. The training loss for all three patients is given in Figure \ref{fig:real_train}(A). For training on data from a single patient, we excluded the adversarial loss from the total loss. As seen in the figure, our network converged within 50 epoch ($\approx 4500$ iteration) on every patient. It took 10 minutes on Nvidia 2080 Ti GPU to train our network. As mentioned above, our method provides a practical solution for the clinical environment, where within a few minutes visibility of an LDCT scans can be improved. One example of the network's denoising performance is given in Figure \ref{fig:real_train}(B). Here we can see the visibility of severely distorted low dose CT image is significantly improved by our method. 
% Please add the following required packages to your document preamble:
% \usepackage{graphicx}
An example of the denoising performance of the proposed method is given in Figure \ref{fig:real_full}. Here we can see denoising performance of BM3D is the worst compared to other methods. On the other hand, the output of the deep image prior is very blurry,  different organ boundary is distorted, and some splotchy artifacts appear in the image. The performance of DIP depends mainly on the stopping iteration; with more iterations, it will again produce the original noisy image. The clean image produced by it is always blurry and without any texture. Consistent with our previous example, we can see CycleGAN has a lot of residual noise left in the example. Next, we identified one hypodense lesion in the image and marked the region with the red colour bounding box. The zoomed version of the region containing the lesion is shown in Figure \ref{fig:real_crop}. The effect of denoising is perceptible in this zoomed view. The visibility of the lesion is very inconspicuous in the original LDCT image. The noise variance is very high in this region; consequently, BM3D and CycleGAN failed to remove noise from this region. On the other hand, DIP has removed the noise but also destroyed the image by removing all the texture information. ConsensusNet also produced a blurry version of the lesion in the output. At the same time, our method produced a denoised image with the lowest granular pattern and improved the lesion's visibility. The main objective of image denoising is to restore the visibility of these types of lesions and anomalies by concealing the noise. In this regard, our method has reached the goal as the perceptibility of different anomalies has been improved vastly without losing structural or textural information. 
\section{Conclusion}
This study proposed a novel method for self-supervised LDCT denoising. The proposed method is built upon the well-celebrated Noise2Noise paradigm. We have proposed an innovative way to produce the second noisy image for training denoiser via taking the average of the two neighbouring slices of each slice. The learnt denoiser is then improved by leveraging the recent concept of invertible learnable blocks and adding a cycle consistency loss in the training stage. The proposed method is an elegant way to use inter-slice correlation for LDCT denoising. Moreover, the proposed method is straightforward to deploy and does not interfere with the workflow of the existing CT scanner. Comprehensive evaluation on both synthetic low dose and real low dose CT data validates our method's superiority over other self-supervised methods.  
{\small
\bibliographystyle{ieee_fullname}
\bibliography{ref}

\begin{thebibliography}{10}\itemsep=-1pt

\bibitem{batson2019noise2self}
Joshua Batson and Loic Royer.
\newblock Noise2self: Blind denoising by self-supervision.
\newblock In {\em International Conference on Machine Learning}, pages
  524--533. PMLR, 2019.

\bibitem{brenner2004radiation}
David~J Brenner.
\newblock Radiation risks potentially associated with low-dose ct screening of
  adult smokers for lung cancer.
\newblock {\em Radiology}, 231(2):440--445, 2004.

\bibitem{chen2017low}
Hu Chen, Yi Zhang, Mannudeep~K Kalra, Feng Lin, Yang Chen, Peixi Liao, Jiliu
  Zhou, and Ge Wang.
\newblock Low-dose ct with a residual encoder-decoder convolutional neural
  network.
\newblock {\em IEEE transactions on medical imaging}, 36(12):2524--2535, 2017.

\bibitem{dabov2006image}
Kostadin Dabov, Alessandro Foi, Vladimir Katkovnik, and Karen Egiazarian.
\newblock Image denoising with block-matching and 3d filtering.
\newblock In {\em Image Processing: Algorithms and Systems, Neural Networks,
  and Machine Learning}, volume 6064, page 606414. International Society for
  Optics and Photonics, 2006.

\bibitem{dinh2014nice}
Laurent Dinh, David Krueger, and Yoshua Bengio.
\newblock Nice: Non-linear independent components estimation.
\newblock {\em arXiv preprint arXiv:1410.8516}, 2014.

\bibitem{dinh2016density}
Laurent Dinh, Jascha Sohl-Dickstein, and Samy Bengio.
\newblock Density estimation using real nvp.
\newblock {\em arXiv preprint arXiv:1605.08803}, 2016.

\bibitem{heusel2017gans}
Martin Heusel, Hubert Ramsauer, Thomas Unterthiner, Bernhard Nessler, and Sepp
  Hochreiter.
\newblock Gans trained by a two time-scale update rule converge to a local nash
  equilibrium.
\newblock {\em Advances in neural information processing systems}, 30, 2017.

\bibitem{huang2017densely}
Gao Huang, Zhuang Liu, Laurens Van Der~Maaten, and Kilian~Q Weinberger.
\newblock Densely connected convolutional networks.
\newblock In {\em Proceedings of the IEEE conference on computer vision and
  pattern recognition}, pages 4700--4708, 2017.

\bibitem{huang2020cagan}
Zhiyuan Huang, Zixiang Chen, Qiyang Zhang, Guotao Quan, Min Ji, Chengjin Zhang,
  Yongfeng Yang, Xin Liu, Dong Liang, Hairong Zheng, et~al.
\newblock Cagan: A cycle-consistent generative adversarial network with
  attention for low-dose ct imaging.
\newblock {\em IEEE Transactions on Computational Imaging}, 6:1203--1218, 2020.

\bibitem{huber1992robust}
Peter~J Huber.
\newblock Robust estimation of a location parameter.
\newblock In {\em Breakthroughs in statistics}, pages 492--518. Springer, 1992.

\bibitem{johnson2016perceptual}
Justin Johnson, Alexandre Alahi, and Li Fei-Fei.
\newblock Perceptual losses for real-time style transfer and super-resolution.
\newblock In {\em European conference on computer vision}, pages 694--711.
  Springer, 2016.

\bibitem{kingma2018glow}
Diederik~P Kingma and Prafulla Dhariwal.
\newblock Glow: Generative flow with invertible 1x1 convolutions.
\newblock {\em arXiv preprint arXiv:1807.03039}, 2018.

\bibitem{krull2019noise2void}
Alexander Krull, Tim-Oliver Buchholz, and Florian Jug.
\newblock Noise2void-learning denoising from single noisy images.
\newblock In {\em Proceedings of the IEEE/CVF Conference on Computer Vision and
  Pattern Recognition}, pages 2129--2137, 2019.

\bibitem{kwon2021cycle}
Taesung Kwon and Jong~Chul Ye.
\newblock Cycle-free cyclegan using invertible generator for unsupervised
  low-dose ct denoising.
\newblock {\em arXiv preprint arXiv:2104.08538}, 2021.

\bibitem{laine2019high}
Samuli Laine, Tero Karras, Jaakko Lehtinen, and Timo Aila.
\newblock High-quality self-supervised deep image denoising.
\newblock {\em arXiv preprint arXiv:1901.10277}, 2019.

\bibitem{lehtinen2018noise2noise}
Jaakko Lehtinen, Jacob Munkberg, Jon Hasselgren, Samuli Laine, Tero Karras,
  Miika Aittala, and Timo Aila.
\newblock Noise2noise: Learning image restoration without clean data.
\newblock {\em arXiv preprint arXiv:1803.04189}, 2018.

\bibitem{9059965}
Zeheng Li, Junzhou Huang, Lifeng Yu, Yujie Chi, and Mingwu Jin.
\newblock Low-dose ct image denoising using cycle-consistent adversarial
  networks.
\newblock In {\em 2019 IEEE Nuclear Science Symposium and Medical Imaging
  Conference (NSS/MIC)}, pages 1--3, 2019.

\bibitem{li2020investigation}
Zeheng Li, Shiwei Zhou, Junzhou Huang, Lifeng Yu, and Mingwu Jin.
\newblock Investigation of low-dose ct image denoising using unpaired deep
  learning methods.
\newblock {\em IEEE Transactions on Radiation and Plasma Medical Sciences},
  5(2):224--234, 2020.

\bibitem{liang2021hierarchical}
Jingyun Liang, Andreas Lugmayr, Kai Zhang, Martin Danelljan, Luc Van~Gool, and
  Radu Timofte.
\newblock Hierarchical conditional flow: A unified framework for image
  super-resolution and image rescaling.
\newblock In {\em Proceedings of the IEEE/CVF International Conference on
  Computer Vision}, pages 4076--4085, 2021.

\bibitem{liu2021invertible}
Yang Liu, Zhenyue Qin, Saeed Anwar, Pan Ji, Dongwoo Kim, Sabrina Caldwell, and
  Tom Gedeon.
\newblock Invertible denoising network: A light solution for real noise
  removal.
\newblock In {\em Proceedings of the IEEE/CVF Conference on Computer Vision and
  Pattern Recognition}, pages 13365--13374, 2021.

\bibitem{sajjadi2017enhancenet}
Mehdi~SM Sajjadi, Bernhard Scholkopf, and Michael Hirsch.
\newblock Enhancenet: Single image super-resolution through automated texture
  synthesis.
\newblock In {\em Proceedings of the IEEE International Conference on Computer
  Vision}, pages 4491--4500, 2017.

\bibitem{su2019low}
Alvin~W Su, Travis~J Hillen, Eric~P Eutsler, Asheesh Bedi, James~R Ross,
  Christopher~M Larson, John~C Clohisy, and Jeffrey~J Nepple.
\newblock Low-dose computed tomography reduces radiation exposure by 90\%
  compared with traditional computed tomography among patients undergoing
  hip-preservation surgery.
\newblock {\em Arthroscopy: The Journal of Arthroscopic \& Related Surgery},
  35(5):1385--1392, 2019.

\bibitem{ulyanov2018deep}
Dmitry Ulyanov, Andrea Vedaldi, and Victor Lempitsky.
\newblock Deep image prior.
\newblock In {\em Proceedings of the IEEE conference on computer vision and
  pattern recognition}, pages 9446--9454, 2018.

\bibitem{van2019reversible}
Tycho~FA van~der Ouderaa and Daniel~E Worrall.
\newblock Reversible gans for memory-efficient image-to-image translation.
\newblock In {\em Proceedings of the IEEE/CVF Conference on Computer Vision and
  Pattern Recognition}, pages 4720--4728, 2019.

\bibitem{wu2019consensus}
Dufan Wu, Kuang Gong, Kyungsang Kim, Xiang Li, and Quanzheng Li.
\newblock Consensus neural network for medical imaging denoising with only
  noisy training samples.
\newblock In {\em International Conference on Medical Image Computing and
  Computer-Assisted Intervention}, pages 741--749. Springer, 2019.

\bibitem{xiao2020invertible}
Mingqing Xiao, Shuxin Zheng, Chang Liu, Yaolong Wang, Di He, Guolin Ke, Jiang
  Bian, Zhouchen Lin, and Tie-Yan Liu.
\newblock Invertible image rescaling.
\newblock In {\em European Conference on Computer Vision}, pages 126--144.
  Springer, 2020.

\bibitem{xuan2004reversible}
Guorong Xuan, Chengyun Yang, Yizhan Zhen, Yun~Q Shi, and Zhicheng Ni.
\newblock Reversible data hiding based on wavelet spread spectrum.
\newblock In {\em IEEE 6th Workshop on Multimedia Signal Processing, 2004.},
  pages 211--214. IEEE, 2004.

\bibitem{zhao2021invertible}
Rui Zhao, Tianshan Liu, Jun Xiao, Daniel~PK Lun, and Kin-Man Lam.
\newblock Invertible image decolorization.
\newblock {\em IEEE Transactions on Image Processing}, 30:6081--6095, 2021.

\bibitem{zhu2017unpaired}
Jun-Yan Zhu, Taesung Park, Phillip Isola, and Alexei~A Efros.
\newblock Unpaired image-to-image translation using cycle-consistent
  adversarial networks.
\newblock In {\em Proceedings of the IEEE international conference on computer
  vision}, pages 2223--2232, 2017.

\end{thebibliography}
}

\end{document}